\documentclass[10pt]{article}
\usepackage[scale=.78]{geometry}
\usepackage[affil-it]{authblk}
\usepackage[title]{appendix}
\usepackage{booktabs}

\usepackage{hyperref}
\usepackage{amsmath}
\usepackage{amsfonts}
\usepackage{amsthm}
\usepackage{amssymb}
\usepackage{xfrac}
\usepackage{graphicx}
\usepackage{siunitx}
\usepackage{comment}

\title{TumorTwin: A python framework for patient-specific digital twins in oncology}

\author[1]{Michael Kapteyn}
\author[1]{Anirban Chaudhuri}
\author[1,3]{Ernesto A. B. F. Lima}
\author[1]{Graham Pash}
\author[1]{Rafael Bravo}
\author[1]{Karen Willcox}
\author[1,2,4,5,6]{Thomas E. Yankeelov}
\author[1,5]{David A. Hormuth II}

 \affil[1]{\footnotesize Oden Institute for Computational Engineering and Sciences, The University of Texas at Austin, TX, USA}
 \affil[2]{\footnotesize Department of Biomedical Engineering, The University of Texas at Austin, Austin, TX, USA}
 \affil[3]{\footnotesize Texas Advanced Computing Center, The University of Texas at Austin, Austin, TX, USA}
\affil[4]{\footnotesize Department of Diagnostic Medicine, The University of Texas at Austin, Austin, TX, USA}
\affil[5]{\footnotesize Livestrong Cancer Institutes, The University of Texas at Austin, Austin, TX, USA}
\affil[6]{\footnotesize Department of Imaging Physics, MD Anderson Cancer Center, Houston, TX, USA}

\newcommand{\bs}{\boldsymbol}
\newcommand{\figsize}{0.45\linewidth}

\date{\vspace{-3em}}

\begin{document}

\maketitle

\begin{abstract}
\noindent\textbf{Background:} Advances in the theory and methods of computational oncology have enabled accurate characterization and prediction of tumor growth and treatment response on a patient-specific basis. This capability can be integrated into a digital twin framework in which bi-directional data-flow between the physical tumor and the digital tumor facilitate dynamic model re-calibration, uncertainty quantification, and clinical decision-support via recommendation of optimal therapeutic interventions. However, many digital twin frameworks rely on bespoke implementations tailored to each disease site, modeling choice, and algorithmic implementation.\\
\textbf{Findings:} We present \texttt{TumorTwin}, a modular software framework for initializing, updating, and leveraging patient-specific cancer tumor digital twins. \texttt{TumorTwin} is publicly available as a Python package, with associated documentation, datasets, and tutorials. Novel contributions include the development of a patient-data structure adaptable to different disease sites, a modular architecture to enable the composition of different data, model, solver, and optimization objects, and CPU or GPU parallelized implementations of forward model solves and gradient computations. We demonstrate the functionality of \texttt{TumorTwin} via an \textit{in silico} dataset of high-grade glioma growth and response to radiation therapy.\\
\textbf{Conclusions:}  The \texttt{TumorTwin} framework enables rapid prototyping and testing of image-guided oncology digital twins. This allows researchers to systematically investigate different models, algorithms, disease sites, or treatment decisions while leveraging robust numerical and computational infrastructure. 
\end{abstract}


\section{Background}
The recent advancement of digital twin (DT) technologies for biomedical applications ~\cite{nasem_dt_2023,laubenbacher2024digital,niarakis2024immune,sel2025survey}, in general, and oncology~\cite{wu_integrating_2022,stahlberg_exploring_2022,chaudhuri_dt_2023}, in particular, has the potential to transform personalized medicine by enabling accurate predictions of disease progression and treatment response as well as identifying improved therapeutic interventions~\cite{Zahid2021,LEDER2014603}. However, many DTs use custom numerical or computational methods, require proprietary code bases, and are tailored for each specific application resulting in limited portability across disease sites and application domains. While we advocate that DTs should be fit-for-purpose and tailored to their domain of application, there is substantial overlap in the data processing, modeling, and numerical solver infrastructure that can be generalized across disease sites to improve accessibility of DT technologies. To this end, we have developed an open-source software package that implements a DT framework with numerical and computational methods designed to facilitate the research and development of DT formulations in oncology.

At the core of a predictive DT in oncology is the computational model that represents a specific patient's tumor and is used to forecast disease progression and response to therapy. However, given the complex multidisciplinary and multiscale nature of cancer, tumor growth modeling is an active area of research. These models often combine first-principles biology and physics with data-driven or phenomenological modeling. Many open questions remain about specific modeling choices, such as \emph{which} effects to model, \emph{how} to model these effects, and how these choices \emph{impact} model accuracy and computational cost~\cite{lima2017selection, kutuva2023}. Additionally, using medical imaging data to calibrate DT models requires complex data processing pipelines (e.g., image registration and segmentation), and the impact of these algorithms on model quality is under-examined~\cite{jarrett_quantitative_2021,bakas2018identifying}. Furthermore, formulation and numerical implementation of calibration and solution algorithms can impact the stability and quality of model predictions~\cite{ghattas2021learning}. A thorough exploration of these tradeoffs requires a modular and adaptable modeling framework so that researchers can establish a baseline end-to-end DT architecture, and then easily experiment with various data, modeling, and algorithmic choices.

Recognizing the need for an adaptable framework to support research in DTs for oncology, we have developed \texttt{TumorTwin}, a python framework for image-guided tumor modeling that supports different data sources, tumor growth models, treatment modules, model parameterizations, numerical solvers, and numerical optimizers. As a baseline, we provide a DT architecture consisting of a well-established tumor growth and treatment model, a suite of performant numerical solvers that support efficient gradient computation, and a suite of numerical optimizers for performing deterministic model calibration. While there exist other frameworks that facilitate computational modeling of cancer, (eg. HAL\cite{bravo2020hybrid}, Chaste\cite{mirams2013chaste}, Physicell\cite{ghaffarizadeh2018physicell}, Netlogo\cite{tisue2004netlogo}, CellSys\cite{abbasi2022cellsys}, Compucell 3D\cite{swat2012multi}, and Morpheus\cite{starruss2014morpheus}), our framework is the first that is specifically designed as an end-to-end (i.e., data-to-decisions) DT framework leveraging tissue-scale imaging data. In particular, creating a DT with a modeling-only library would require the user to implement separate algorithms for data processing and model calibration; we provide a complete high-performance pipeline for this purpose. We demonstrate the use of this framework for creating a DT of a high-grade glioma (HGG)~\cite{HormuthII2021} with a second demonstration focused on triple negative breast cancer~\cite{jarrett_quantitative_2021} provided in the appendix. For both cases, the DT is initialized for an individual patient using their quantitative magnetic resonance imaging (MRI) data to generate a personalized tumor growth model\cite{jarrett_quantitative_2021,hormuth2021image}. Future MRI visits can be used to further calibrate model parameters related to tumor growth and response to treatment. The calibrated DT can be used to predict growth under different candidate treatments, which in turn can be used to optimize treatment on a patient-specific basis.

This paper presents the \texttt{TumorTwin} package, providing a discussion on the underlying theory, design principles, and implementation details. We also provide a demonstrative case-study using an \textit{in silico} HGG patient, intended to showcase the package functionality and provide a performance analysis for a representative case.

\section{Findings}
\subsection{Design principles and overview}
The goal of this software package is to empower researchers to develop high-performance predictive DTs for oncology, leveraging medical imaging datasets and incorporating treatment protocols. A key focus of our design is to balance performance with usability, ensuring that researchers can easily modify and extend various aspects of the data pipeline, computational models, and solvers. This allows researchers to explore new modeling directions and computational technologies in the context of an end-to-end DT workflow.

To achieve this flexibility, we have developed a modular codebase with well-defined abstractions. This modularity enables researchers to swap, customize, or extend different components without requiring deep modifications to the core framework. Each module is designed to interact seamlessly with others, facilitating an intuitive workflow for building and refining patient-specific DTs. This standardized workflow guides users through the key steps of building a predictive model:
\begin{enumerate}
\item Prepare input data (e.g., imaging, treatment history).
\item Construct a \texttt{PatientData} object to encapsulate all relevant patient-specific information.
\item Generate a \texttt{Model} object based on this patient data, encoding tumor growth and treatment response dynamics.
\item Wrap the model in a \texttt{Solver}, which performs numerical integration or simulations to generate predictions.
\item Optionally use an \texttt{Optimizer}, which refines model parameters to best match observed data.
\end{enumerate}
This workflow ensures clarity and reproducibility while allowing researchers to incorporate custom components at each stage. For example, the \texttt{Model} can be expanded to include new treatment terms or biological features (e.g., mechanics-coupled tumor growth~\cite{Hormuth2017}) Figure 1 provides a high-level overview of the package structure, illustrating how these components interact. Each element is described in detail in the following sections. 
\begin{figure}[!htb]
    \label{fig:software-architecture-workflow}
    \centering
    \includegraphics[width=\figsize]{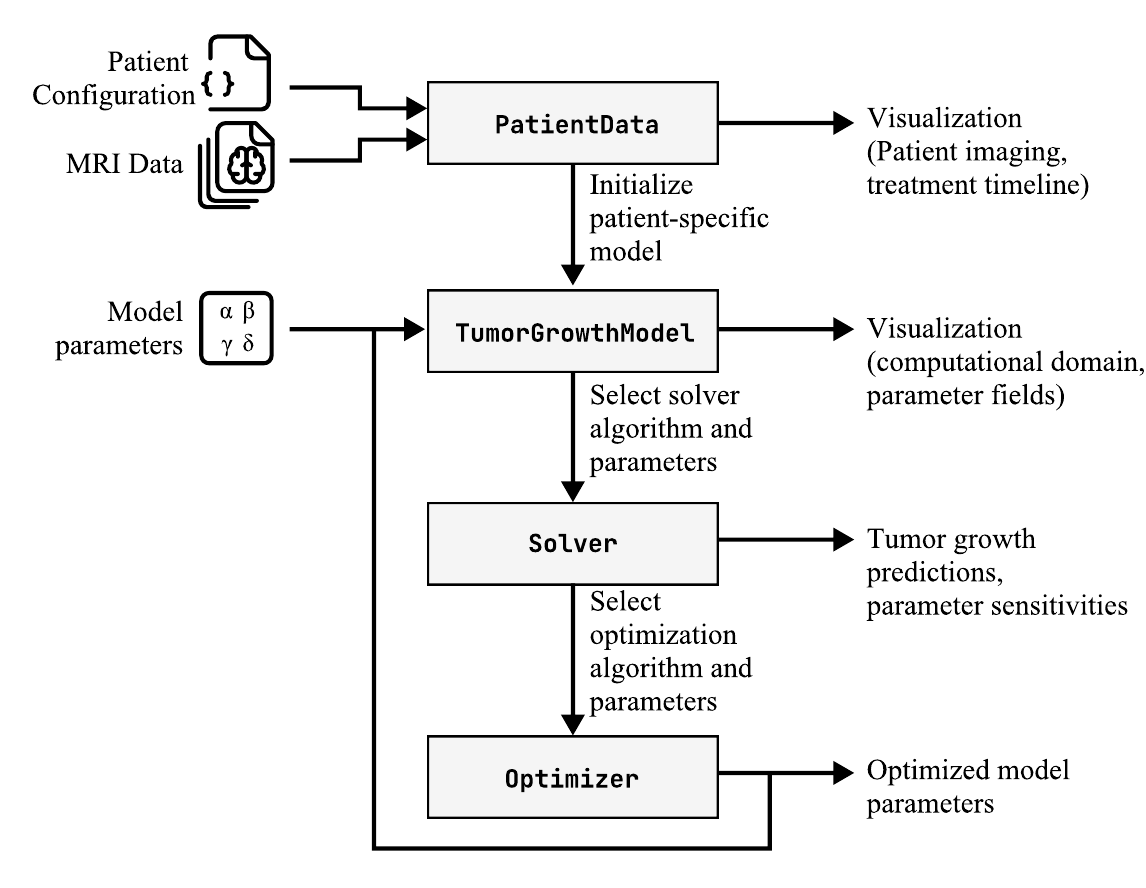}
    \caption{\textit{{\normalfont \texttt{TumorTwin}} workflow key components.} \texttt{TumorTwin} takes as inputs the patient configuration file, MRI data, and an initial guess of model parameters. The patient configuration file and MRI data are used to construct a \texttt{PatientData} object which serves as the central data object throughout \texttt{TumorTwin}. The \texttt{PatientData} object is used to initialize a patient-specific \texttt{TumorGrowthModel}, which can be combined with a \texttt{ForwardSolver} to simulate tumor growth over time. If longitudinal data is available within the \texttt{PatientData} object, an \texttt{Optimizer} module can be used to calibrate model parameters by minimizing the error between model predictions and patient-specific measurements.
    }
\end{figure}

The package is written entirely in Python and leverages \texttt{PyTorch}, providing several advantages:
\begin{enumerate}
\item GPU compatibility, allowing computationally intensive tumor simulations to run efficiently on either CPU- or GPU-based computing platforms.
\item Automatic differentiation, enabling sensitivity analysis and seamless integration of gradient-based optimization techniques for parameter fitting.
\item Extensibility, as users can compose our models with other PyTorch-based architectures, such as neural ODEs, learned pre-processing pipelines, or deep-learning-based post-processing steps.
\end{enumerate}

\subsection{Implementation}
\subsubsection{Example datasets provided with the package}
The primary data source used in our framework is medical imaging data, which serves as the foundation for constructing and calibrating patient-specific DT models. However, working with medical imaging data presents several challenges, e.g., large file sizes and complex file formats with multiple conventions for coordinate systems, units, and other metadata. These factors make preprocessing and standardization critical steps in any DT pipeline.

In addition to these technical challenges, the use of real patient imaging data is further complicated by patient privacy concerns. Sharing clinical imaging datasets openly typically requires extensive de-identification procedures and complex institutional approval processes. As a result, publicly available datasets that can be used for testing and benchmarking are often limited in scope and accessibility.

To provide users with accessible demonstrations and a reference for dataset creation, we have developed and included two datasets that have been synthesized using real patient data as a reference:
\begin{enumerate}
    \item A dataset for high-grade glioma (HGG), featuring synthetic brain MRI scans and corresponding radiotherapy and chemotherapy schedules.  
    \item A dataset for triple-negative breast cancer (TNBC), containing synthetic breast imaging data and chemotherapy schedules.
\end{enumerate}  
These datasets are designed to facilitate quick exploration of the software’s capabilities without requiring access to real clinical data, while also serving as structured templates for users who wish to integrate their own patient datasets into the framework. The procedure for generating these synthetic datasets is detailed in the Methods section. Throughout this report, we demonstrate the functionality of our package using the HGG dataset. To illustrate the generalizability of the approach across different cancer types, we provide an analogous TNBC demonstration in \hyperref[sec:appendix]{\textit{Appendix A}}.

\subsubsection{Importing and pre-processing patient data}
A critical first step in constructing a DT is importing patient-specific data, which typically includes medical imaging and treatment history. Our package is designed to handle these inputs efficiently while ensuring interoperability with existing medical imaging and data processing tools.  

To facilitate integration with existing medical imaging workflows, our package supports the \textit{Neuroimaging Informatics Technology Initiative} (NIFTI) format, a widely used format for medical imaging data. We provide a lightweight wrapper around the NIFTI classes from established Python libraries, including \texttt{nibabel} and \texttt{ITK/SimpleITK}. This approach ensures full interoperability with these widely adopted toolkits, allowing users to leverage their extensive functionality for image processing, registration, and analysis while using our package for DT modeling.

Ensuring the integrity and compatibility of patient data is essential for reliable model predictions. To this end, we use the \texttt{Pydantic} data validation library to define a structured data model with built-in validations. We implement a \texttt{BasePatientData} class, which defines a base data model comprised of one-off imaging data (e.g. anatomic masks), longitudinal treatment data, and a list of visits, each associated with a time and visit-specific imaging dataset. We implement specialized subclasses for different cancer types, e.g.,the \texttt{HGGPatientData} class specifies the imaging formats required, for a HGG DT. For the \texttt{HGGPatientData} class, standard anatomical  ($T_1$-weighted and $T_2$-fluid attenuated inversion recovery (FLAIR)) and functional (apparent diffusion coefficient, $ADC$, from diffusion-weighted MRI) are required.

Based off previous studies~\cite{jarrett_quantitative_2021,hormuth2021image},  $ADC$ is used to assign the observational data $\bs{o}=\left[o(t_1)^\top,\dots,o(t_{n_\text{visit}})^\top \right]$ for $n_\text{visit}$ number of visits. Using the \texttt{ADC\_to\_cellularity} function, the observation at each visit $\left\{t_i\right\}_{i=1}^{n_\text{visit}}$ is defined as
\begin{equation}\label{eqn:adc}
    o(t_i) = N(\bs{x},t_i) = \frac{ADC_\text{w}-ADC(\bs{x},t_i)}{ADC(\bs{x},t_i)-ADC_\text{min}},
\end{equation}
where $N(\bs{x},t_i)$ is the tumor cellularity at 3D position $\bs{x}$ and time $t_i$, $ADC_\text{w}$ is the $ADC$ of water at room temperature ($3.0 \times 10^{-3}\ \text{mm}^2/\text{s}$~\cite{Woodhams2011}), $ADC_\text{min}$ is the minimum observed $ADC$ within the tumor region of interest.  The \texttt{ADC\_to\_cellularity} function only assigns $N$ within the tumor regions of interest, and assigned zero-elsewhere.

To provide flexibility in bundling imaging data across multiple modalities and multiple imaging visits, alongside patient treatment data, we employ JSON-based configuration files which are loaded into the corresponding \texttt{pydantic} patient data object. JSON is a lightweight and platform-agnostic format, which allows users to easily store, share, and version control their DT configuration files. When creating a patient data object (either manually via Python code, or by loading a JSON file), \texttt{Pydantic} validation verifies that the provided MRI data and treatment records are consistent and complete, helping to catch formatting issues early and reducing potential errors in model building and downstream model computations.

To assist in verifying and interpreting input data, we provide a visualization method for \texttt{PatientData} objects. This method generates a summary figure that includes:  
\begin{itemize}
    \item A treatment timeline, displaying administered therapies, doses, and imaging timepoints. 
    \item The provided medical images and regions of interest.  
     
\end{itemize}
These visualizations serve as a quick diagnostic tool, allowing users to inspect patient-specific data before initiating model simulations.  
\begin{figure}
    \centering
    \includegraphics[width=\figsize]{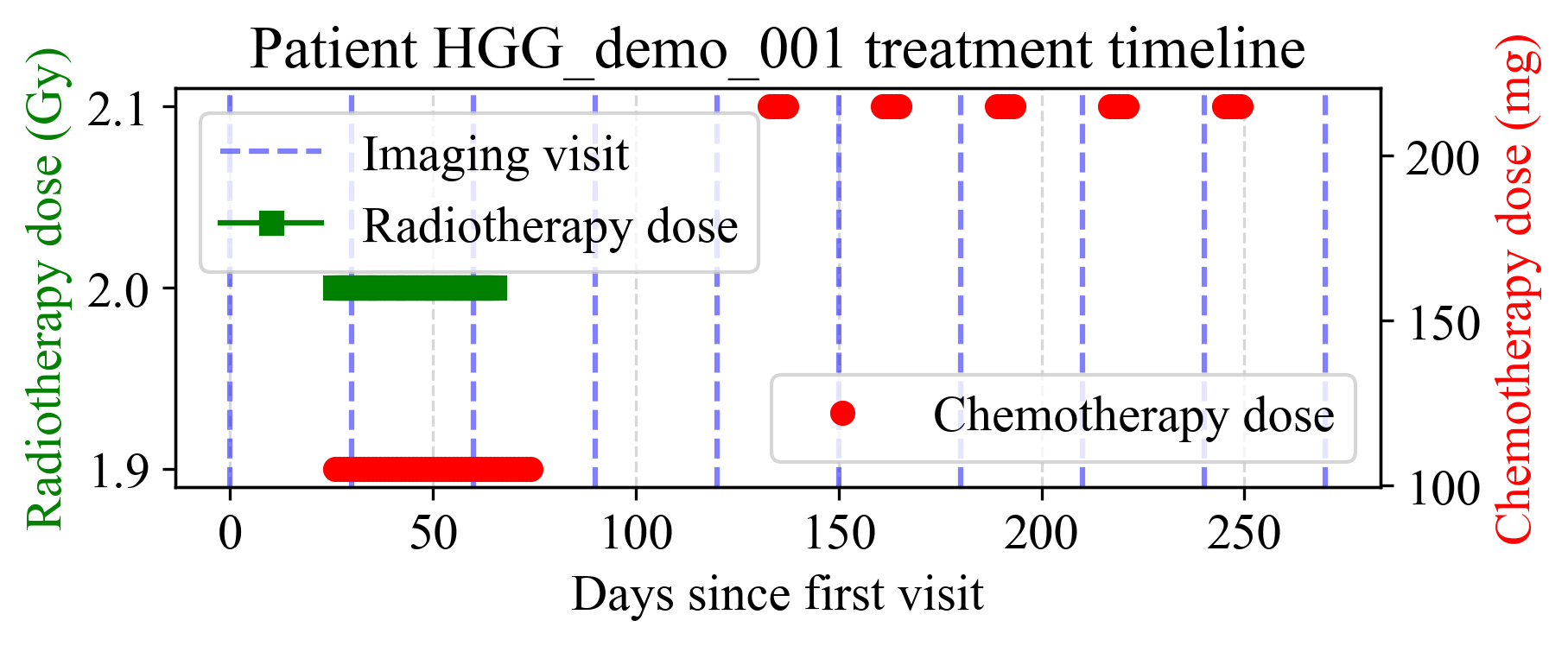}\\ \vspace{-0.5em}
    \includegraphics[width=\figsize]{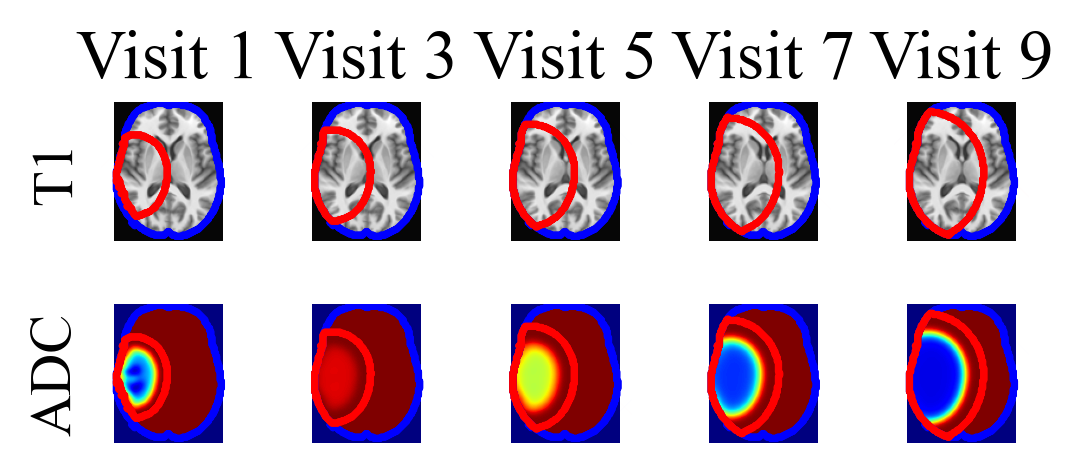}
    \caption{\textit{Patient data summary graphics.} Example output of the patient data summary function applied to the \textit{in silico} HGG dataset. This graphic shows treatment and imaging schedule (top panel), anatomical $T_1$-weighted MRI  with associated tumor segmentations (middle panel), and the apparent diffusion coefficient ($ADC$) map for the same imaging slices (bottom panel).  In practice, this summary can be used to visually confirm longitudinal registration of imaging series, accuracy of tumor segmentations, and treatment details.}
    \label{fig:patient_data_summary}
\end{figure}

\subsubsection{Mathematical model of tumor growth and treatment response}
Our base tumor model class, \texttt{TumorGrowthModel3D}, represents a system of ordinary differential equations, such as those those arising from the spatial discretization of a time-evolving PDE~\cite{quarteroni2008numerical}, in the form:
\begin{eqnarray}\label{eqn:ode}
    \frac{du}{dt} = f(u,t,\bs{p}) \\ 
    u(0) = u_0
\end{eqnarray}
where $u(t)$ represents some characteristic of the tumor, $t$ represents time, $u_0$ is the initial condition, and $\bs{p}$ represents a set of model parameters. The required method in this class is the \texttt{forward} method, which evaluates the right-hand side of Equation~\ref{eqn:ode}. This allows users to implement their own mathematical model by implementing a subclass of the base tumor growth model complete with a corresponding \texttt{forward} method.

On top of this general foundation, we have provided a specific implementation of a commonly used mathematical model~\cite{jarrett_quantitative_2021,HormuthII2021,hormuth_opportunities_2022}, which takes the form of a reaction-diffusion partial differential equation (PDE) with chemotherapy (CT) and radiotherapy (RT) treatment effects. The reaction-diffusion model describes the change in the tumor cellularity due to tumor cell invasion (diffusion), logistic growth (reaction), and death due to treatment (RT and CT):
\begin{align}\label{eqn:RD}
    \frac{\partial N(\bs{x},t)}{\partial t} &= \underbrace{\nabla \cdot \left( D \nabla N(\bs{x},t) \right)}_{\text{invasion}} \notag + \underbrace{k(\bs{x})N(\bs{x},t) \left(1 - \frac{N(\bs{x},t)}{\theta} \right)}_{\text{logistic growth}} \notag \\
    &\quad \underbrace{- \sum_{i=1}^{n_{\text{CT}}} \sum_{j=1}^{T_i}\alpha_i C_i \exp\left(-\beta_i\left( t-\tau_{i,j}\right)\right) N(\bs{x},t)}_{\text{chemotherapy}} ,\\
    N(\bs{x},t)_{\text{after}} &= \underbrace{N(\bs{x},t)_{\text{before}} \exp\left(-\alpha_\text{RT}d_\text{RT}(t)-\beta_\text{RT}d_\text{RT}^2(t)\right)}_{\text{radiotherapy}}   \label{eqn:RT}
\end{align}
where $N$ is the tumor cell density (units: $\text{cells}/\text{mm}^3$), $D$ is the tumor cell diffusion coefficient (units: $\text{mm}^2/\text{day}$),  $k(\bs{x})$ is the tumor cell proliferation rate (units: $\text{day}^{-1}$),  $\theta$ is the carrying capacity (units: \text{cells}), $n_{\text{CT}}$ is the number of different CT agents, $T_i$ is the total number of doses delivered for agent $i$, $\alpha_i$ is the efficacy of CT agent $i$ (units: $\text{day}^{-1}$), $C_i$ is the normalized dose of the CT agent $i$, $\beta_i$ is the decay rate for CT agent \textit{i} (units: $\text{day}^{-1}$), and $\tau_{i,j}$ is the time of $j$-th administration of CT agent $i$. Equation~\eqref{eqn:RT} defines the effect of radiotherapy and is modeled as an instantaneous reduction in the tumor cellularity at the time of delivery, with the survival fraction based on the linear quadratic model~\cite{mcmahon2018linear}. Here $N_{\text{before}}$ and $N_{\text{after}}$ are the tumor cellularity immediately before and after an RT event, $\alpha_\text{RT}$ and  $\beta_\text{RT}$ are radiosensitivity parameters (units: $\text{Gy}^{-1}$ and $\text{Gy}^{-2}$, respectively), and $d_\text{RT}(t)$ is the RT dose delivered at time $t$.

Equations~\eqref{eqn:RD}-\eqref{eqn:RT} form the foundational tumor growth and treatment model \texttt{ReactionDiffusion3D} provided with in \texttt{TumorTwin}. This model serves as a flexible template, enabling the adaptation of existing models or the integration of new ones while ensuring compatibility with the \texttt{Solver}, \texttt{PatientData}, and \texttt{Optimizer} objects. This design facilitates the rapid deployment of novel models within the DT framework.

The model represented by Equation~\eqref{eqn:RD} is spatially discretized using a finite-difference scheme~\cite{leveque2007finite}, with a grid size that corresponds to the voxel size in the input MRI data. This gives rise to a system of coupled ODEs of the form Equation~\eqref{eqn:ode} with discrete radiotherapy events:
\begin{align}
\label{eqn:discretized}
    \frac{d\mathbf{N}}{dt} &= D \mathbb{L} \mathbf{N} + k \mathbf{N} \left(1 - \frac{\mathbf{N}}{\theta} \right) -
    \sum\limits^{n_{\text{CT}}}_{i=1} \sum\limits^{T_i}_{j=1} \alpha_i C_i \exp\left(-\beta_i\left(t-\tau_{i,j}\right)\right) \mathbf{N},\\
    \mathbf{N}_{\text{after}} &= \mathbf{N}_{\text{before}}\exp\left(-\alpha_\text{RT}d_\text{RT}(t)-\beta_\text{RT}d^2_\text{RT}(t)\right)   \label{eqn:RT_discretized}
\end{align}
where $\mathbf{N}$ is vector representing tumor cellularity in the voxels and $\mathbb{L}$ is the discretized Laplace operator, e.g. \textit{via} a second-order central difference scheme. Note that in general $k$ and $D$ may be spatial fields, but we here assume that they are homogeneous in the domain (i.e., $k$ and $D$ are scalars), which allows us to pre-assemble a Laplacian operator independent of $D$. The model parameters are $\bs{p} = \{k, D, \theta, \bs{\alpha}, \bs{\beta}, \alpha_\text{RT}, \beta_\text{RT}\}$, where $\bs{\alpha}=\left[\alpha_1,\dots,\alpha_{n_\text{CT}}\right]$ and $\bs{\beta}=\left[\beta_1,\dots,\beta_{n_\text{CT}}\right]$. The model is initialized by using the observational data from the first patient visit to compute an initial condition $u_0 = \mathbf{N}_0$ using Equation~\eqref{eqn:adc}. A scalar quantity of interest that reflects the combined size and intensity of the tumor is the total tumor cell count (TTC), which can be computed from the tumor cellularity, $\mathbf{N}$, by computing the number of tumor cells in each voxel and summing across all voxels in the computational domain as
\begin{equation}
\text{TTC}(t) = \sum_\mathbf{i} \mathbf{N}_i(t)\theta\text{V}_{\text{voxel}} \label{eqn:ttc}
\end{equation}

\subsubsection{Model prediction via high-performance solvers}
\begin{figure}[!htb]
    \centering
    \includegraphics[width=\figsize]{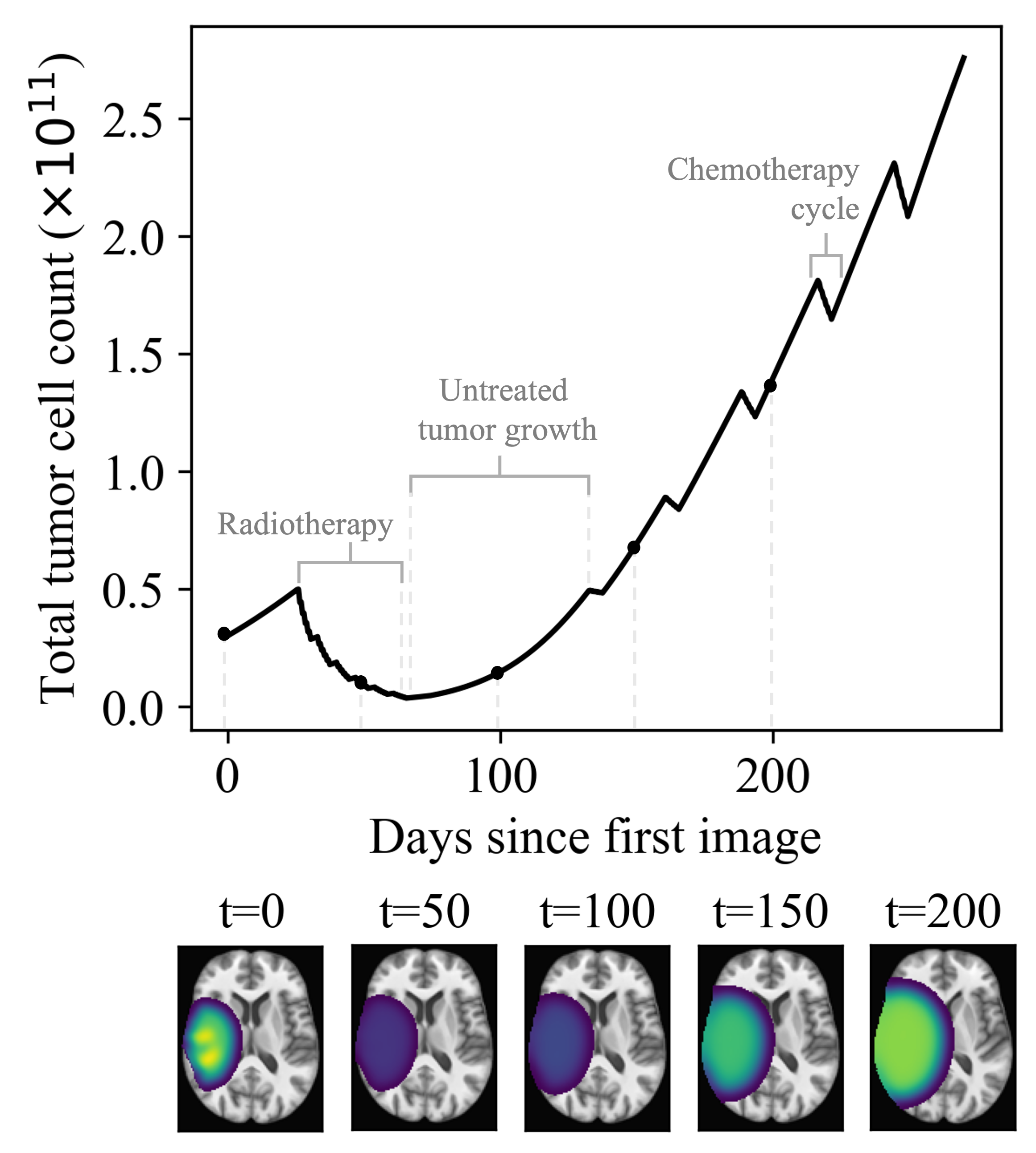}
    \caption{\textit{Example tumor growth prediction for the \textit{in silico} HGG patient}. The predicted total tumor cell count over time is shown (top panel), with 2D slices of the model solution at three snapshots in time (bottom panel). Note the invasion, logistic growth, chemotherapy, and radiotherapy effects in the solution.}  
    \label{fig:prediction}
\end{figure}

Generating tumor growth predictions requires solving the coupled set of ODEs defined by a tumor growth model in the form of Equation~\eqref{eqn:ode}. As the size of the state vector $u$ can be large (e.g., equal to the number of imaging voxels in the computational domain for a spatially-discretized PDE model such as Equation~\eqref{eqn:discretized}) an efficient numerical integration scheme is crucial for tractable simulation. In this work, we employ the \texttt{torchdiffeq} library~\cite{torchdiffeq}, which provides differentiable ODE solvers compatible with the PyTorch framework. Available solver schemes include standard fixed-step schemes such as fourth-order Runge-Kutta (\texttt{rk4})~\cite{hairer1993solving}, as well as adaptive-step methods, such as the fifth-order Runge-Kutta of Dormand-Prince-Shampine (\texttt{dopri5})~\cite{dormand1980family}. The package provides \texttt{TorchDiffEqSolver} that is based on the \texttt{torchdiffeq} library and supports specifying output timesteps independently from the solver timesteps, and the handling of discrete events (for example, to implement the radiotherapy model given by Equation~\eqref{eqn:RT}). As the solver is compatible with PyTorch, forward solves can be run on a GPU architecture simply by setting \texttt{device=torch.device("cuda")}.
Figure~\ref{fig:prediction} shows a characteristic prediction for the model described in the previous section, showing both the TTC (Equation~\eqref{eqn:ttc} over time, and 2D snapshots from the full 3D solution domain at specific timepoints.
 
\subsubsection{Efficient gradient computation}
In addition to solving a model forward in time to predict tumor growth for a given set of model parameters, our package also supports the efficient computation of gradients. When the provided tumor growth model is differentiable and PyTorch compatible, derivatives of output quantities with respect to input parameters can be computed efficiently via reverse-mode automatic differentiation (backpropagation) using the standard \texttt{pytorch} syntax. However, tumor growth models typically require a large number of states (e.g. equal to the number of imaging voxels in the computational domain), and a large number of successive timesteps. Thus, the computational graph that needs to be maintained in order to leverage automatic differentiation often becomes prohibitively large, exceeding the available memory of most systems. An alternative is to use the adjoint method for computing gradients, which requires $\mathcal{O}(1)$ memory, at the cost of requiring a backwards-in-time solve (the adjoint pass) through the model. 

We leverage the adjoint method implemented in the solver library \texttt{torchdiffeq}~\cite{kidger2021hey}, which is available via a simple keyword argument (\texttt{use\_adjoint=True}) in the solver interface. This capability allows one to compute gradients of any scalar function of the solver output, with respect to any of the model inputs. For example, one could run a forward solve over a period of $200$ days to compute $\mathbf{N}(200)$, post-process the solution to compute $\text{TTC}(200)$ via Equation~\eqref{eqn:ttc}, and then run a backward pass to compute $\frac{\partial{(TTC(200))}}{\partial k}$, i.e., the rate of change of the solution with respect to the proliferation rate parameter, $k$.

\subsubsection{Model calibration via gradient-based numerical optimization}
The mathematical model described by Equations~\eqref{eqn:RD}-\eqref{eqn:RT} describes the baseline model that lie at the core of the DT for any patient. To build a DT of a specific patient, one needs to identify the $n_p$ patient-specific parameters $\bs{p}\in \mathcal{P} \subseteq \mathbb{R}^{n_p}$ by calibrating the model to patient-specific observational data $\bs{o}$ (see Equation~\eqref{eqn:adc} and  Figure~\ref{fig:patient_data_summary}). Here we calibrate for $\bs{p} = \{k, D, \alpha_1, \alpha_\text{RT}\}$ and fix the rest of the parameters to their ground truth values (see \textit{Methods} section, \hyperref[sec:data-generation]{\textit{Generation of synthetic longitudinal imaging studies}}). We provide deterministic model calibration capability in the \texttt{TumorTwin} codebase that can identify the patient-specific parameters $\bs{p}^*$ by solving the optimization problem
\begin{equation}
\label{eqn:opt_calibration}
    \bs{p}^* = \underset{\bs{p}\in\mathcal{P}}{\arg\min}\ \mathcal{L}( \bs{p}; \bs{o}),
\end{equation}
where $\mathcal{L}$ is a user-defined scalar loss function that describes the distance between the model predictions and the patient-specific observations for a given value of $\bs{p}$. In this work, we use the mean squared error between predicted cellularity maps and cellularity maps derived from observations as the loss function. Recall that we are able to compute gradients of the loss function with respect to parameters, as described in the previous section. Thus, our solver is compatible with the built-in gradient-based Pytorch optimizers (\texttt{pytorch.optim}) which solve the optimization problem in Equation~\eqref{eqn:opt_calibration}. In addition to the built-in Pytorch optimizers, we have also included a custom implementation of the Levenberg-Marquardt (LM) algorithm for optimization, as LM has been frequently used for patient-specific parameter calibration of PDE models of cancer~\cite{hormuth2021image,hormuth2020forecasting}. The user can easily customize the loss function and optimization algorithm with the rest of the modeling and solver choices, further extending the modularity of the framework.

\begin{figure}[!htb]
    \centering
    \includegraphics[width=\figsize]{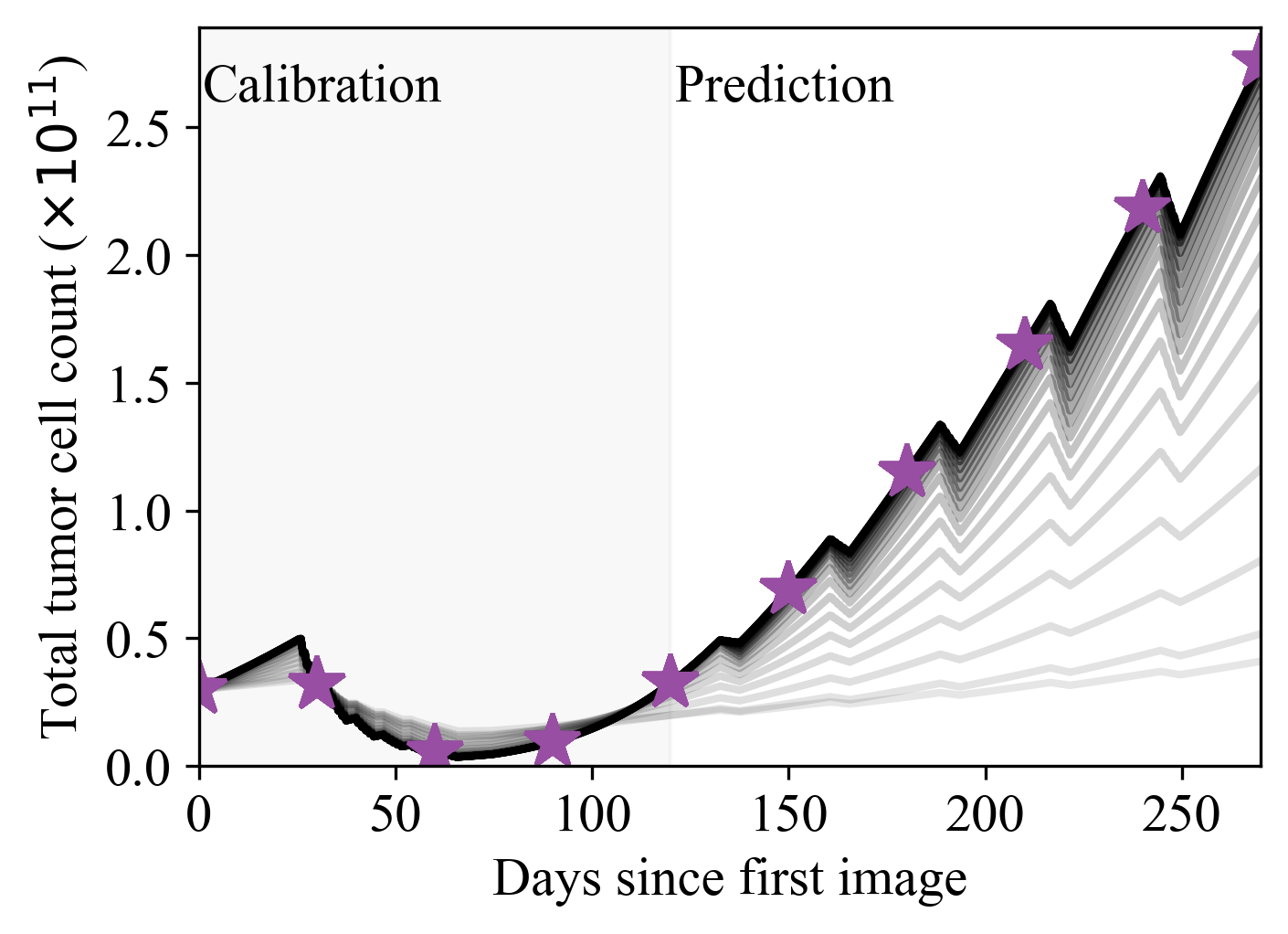}\\ \vspace{-0.5em}
    \includegraphics[width=\figsize]{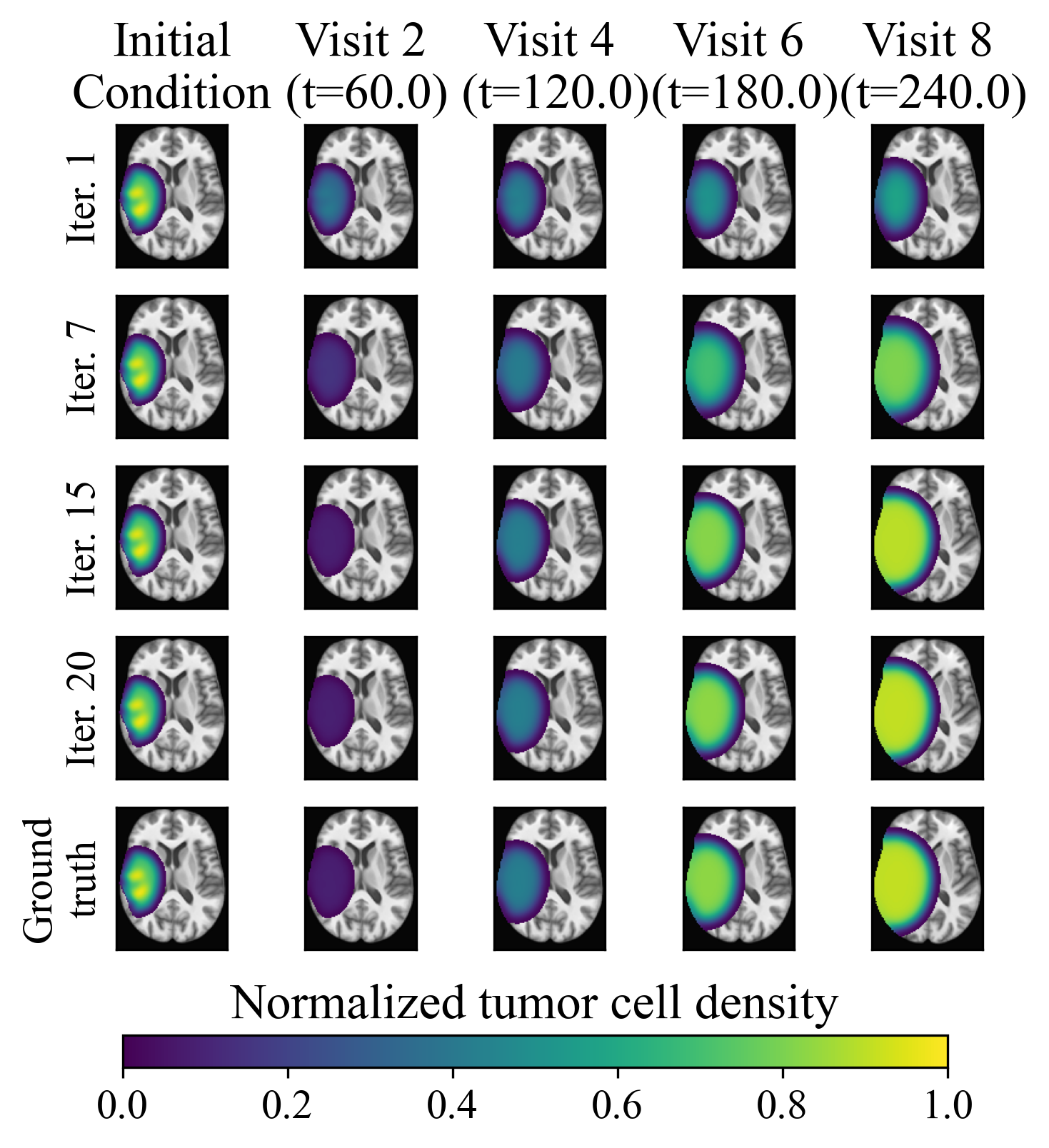}\\ \vspace{-0.5em}
    \includegraphics[width=\figsize]{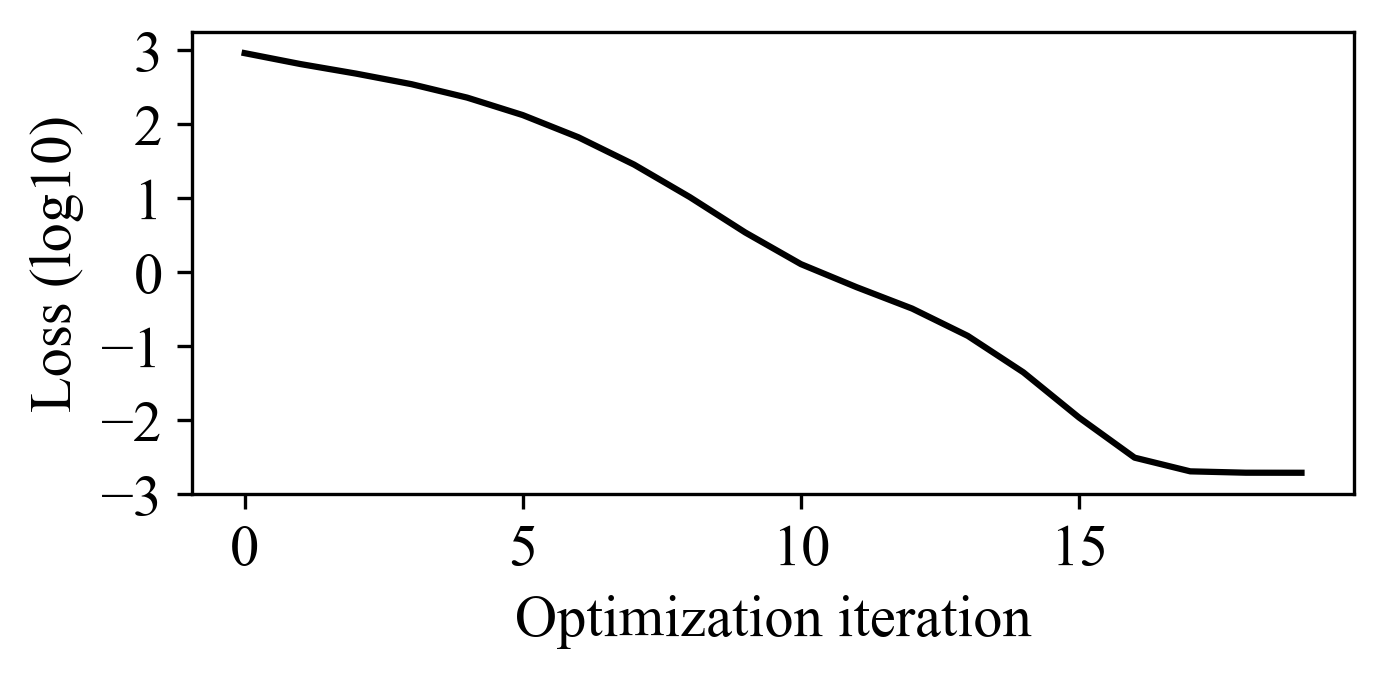}
    \caption{\textit{Model calibration to patient-specific MRI data}. Top: Total tumor cell count (TTC) time series for each iteration of the calibration (black lines; opacity increased with iteration) compared with the observed TTC (purple stars). Middle: Evolution of a central tumor slice across time (left-to-right), and across optimization iterations (top-to-bottom). The bottom row shows the ground-truth data used for calibration and validation. Bottom: Convergence of the loss function over optimization iterations.}
    \label{fig:calibration}
\end{figure}

To demonstrate the model calibration capability, we use the LM optimizer within PyTorch to calibrate a patient-specific DT using the \textit{in silico} HGG dataset. For this example, we know the ground-truth model parameters (used to generate the dataset), but the optimizer is initiated with an initial guess of $k=0.01$, $D=0.02$, $\alpha_1 = 0.01$, $\alpha_\text{RT}=0.04$ (20\% of the respective truth values). Other parameters are fixed to their ground truth values. Figure~\ref{fig:calibration} shows the results of running the optimizer with default optimizer parameter values. For this study, we calibrate the model to the first five imaging timepoints, while holding out the remaining images to assess predictive accuracy. We observe that the optimizer is able to calibrate the unknown model parameters to match the input MRI data within a few iterations, achieving between $10^{-4}$ and $10^{-6}$ relative error in parameter values. We also observe that the solver and optimizer are robust to large variations in the solution across iterations. 

Here we have shown the predictive performance of the calibrated tumor growth model for a single treatment schedule. Note that, once calibrated, the model could be used to predict tumor growth and response to \textit{alternative} treatment schedules, simply by changing the treatment parameters (dosage, timing) for treatments in the prediction regime. In this way, the calibrated model can form the basis of a patient-specific DT that could be used to issue predictions of future tumor growth, guide treatment decisions, and/or be re-calibrated whenever more MRI data are acquired. 

\subsubsection{Empirical performance evaluation}
\begin{figure}[!htb]
    \centering
    \includegraphics[width=\figsize]{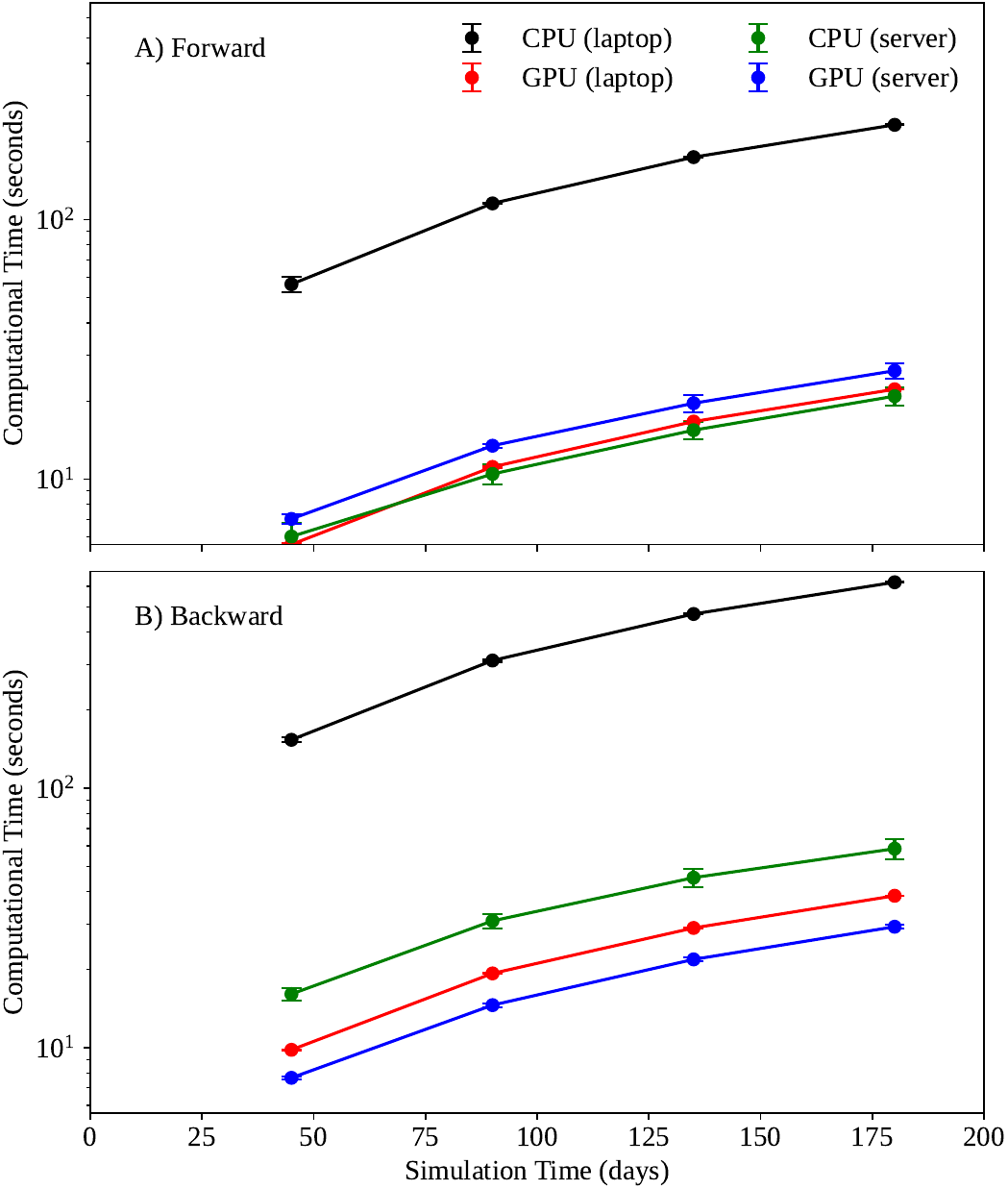}
    \caption{\textit{Solver performance profiling for CPU and GPU architectures.} Top: Mean and standard deviation of the wall-clock time required to compute a forward model prediction for 180 days. Bottom: Results for a backward adjoint solve for the same period. Two CPU-based and two GPU-based hardware architectures were tested.}
    \label{fig:solver_performance}
\end{figure}

We demonstrate the performance of our framework by comparing execution time on CPU vs.\ GPU (CUDA) architectures while varying the length of the tumor simulation between 45, 90, 135, and 180 days. The computational domain is a uniform 3D imaging grid of size ($166 \times 245 \times 48$) voxels, corresponding to a physical domain size of approximately (77.8 mm $\times$ 114.9 mm $\times$ 144.0 mm). This leads to a total of 1,952,160 degrees of freedom in the spatially discretized model. We solve the system of coupled ODEs given by Equation~\eqref{eqn:discretized}, which governs tumor growth and treatment response dynamics. Figure~\ref{fig:solver_performance} and Table~\ref{tab:solver_performance} present the timing results for forward and backward passes across the range of prediction lengths. 

All experiments were conducted on two different systems, representative of computational resources typically available to researchers: a Dell Inspiron 16 Plus 7630 laptop running Ubuntu 22.04.4 LTS, equipped with an Intel Core i7-13700H processor (14 cores, 20 threads, 5.0 GHz max clock), 32 GB RAM, and an NVIDIA GeForce RTX 4060 Laptop GPU (8 GB VRAM, CUDA 12.2); and a Dell PowerEdge R740 server running Ubuntu 22.04.5 LTS, equipped with an Intel Xeon Gold 6248R processor (48 cores, 96 threads, 4.0 GHz max clock), 187 GB RAM, and an NVIDIA A100 PCIe GPU (40 GB VRAM, CUDA 12.4).

Our results show that forward solves scale roughly linearly with the prediction length. On the laptop hardware, we observe an approximately $10\times$ speedup running on the GPU vs. CPU, while backward solves (gradient-based calibration) are more computationally expensive but benefit from an approximately $15\times$ GPU speedup. This acceleration is particularly important when performing deterministic calibration, where many iterations of forward and backward simulations may be required. On the high-performance server architecture we observed significant accelerations in CPU run time vs. the laptop architecture, owing to the more capable hardware and significantly increased memory. Overall, these results suggest that our framework enables the prediction of tumor growth over a period of one year in roughly one minute, with a backward pass in roughly $1-2$ minutes, on either a consumer grade GPU or a high-performance CPU.
\begin{table}[!htb]
    \centering
    \caption{Comparison of solver execution time on CPU vs. GPU for forward (F) and backward (B) solver steps on a laptop and a server. The table shows the mean computational time (in seconds) for each solver step, along with the corresponding standard deviation. Speedup is calculated as the ratio of CPU to GPU time.}
    \label{tab:solver_performance}
    \begin{tabular}{lccc}
        \toprule
        \textbf{Solver Step} & \textbf{CPU Time (s)} & \textbf{GPU Time (s)} & \textbf{Speedup} \\
        \midrule
        \multicolumn{4}{c}{Laptop} \\
        \midrule
    F (45 days) & 56.36 $\pm$ 3.99 & 5.60 $\pm$ 0.05 & \textbf{10.07 $\pm$ 0.72} \\
F (90 days) & 115.50 $\pm$ 0.58 & 11.14 $\pm$ 0.08 & \textbf{10.37 $\pm$ 0.09} \\
F (135 days) & 174.11 $\pm$ 0.56 & 16.66 $\pm$ 0.08 & \textbf{10.45 $\pm$ 0.06} \\
F (180 days) & 231.97 $\pm$ 1.15 & 22.19 $\pm$ 0.07 & \textbf{10.45 $\pm$ 0.06} \\
B (45 days) & 153.75 $\pm$ 3.34 & 9.81 $\pm$ 0.03 & \textbf{15.67 $\pm$ 0.34} \\
B (90 days) & 310.74 $\pm$ 3.60 & 19.34 $\pm$ 0.07 & \textbf{16.07 $\pm$ 0.20} \\
B (135 days) & 469.64 $\pm$ 1.63 & 28.95 $\pm$ 0.15 & \textbf{16.22 $\pm$ 0.10} \\
B (180 days) & 622.31 $\pm$ 1.98 & 38.51 $\pm$ 0.03 & \textbf{16.16 $\pm$ 0.05} \\
        \midrule
        \multicolumn{4}{c}{Server} \\
        \midrule
    F (45 days) & 6.01 $\pm$ 0.75 & 7.02 $\pm$ 0.28 & \textbf{0.86 $\pm$ 0.11} \\
F (90 days) & 10.46 $\pm$ 0.91 & 13.45 $\pm$ 0.24 & \textbf{0.78 $\pm$ 0.07} \\
F (135 days) & 15.43 $\pm$ 1.19 & 19.61 $\pm$ 1.52 & \textbf{0.79 $\pm$ 0.09} \\
F (180 days) & 20.87 $\pm$ 1.66 & 26.16 $\pm$ 1.84 & \textbf{0.80 $\pm$ 0.08} \\
B (45 days) & 16.10 $\pm$ 0.88 & 7.66 $\pm$ 0.10 & \textbf{2.10 $\pm$ 0.12} \\
B (90 days) & 30.82 $\pm$ 1.95 & 14.59 $\pm$ 0.25 & \textbf{2.11 $\pm$ 0.14} \\
B (135 days) & 45.23 $\pm$ 3.78 & 21.90 $\pm$ 0.35 & \textbf{2.07 $\pm$ 0.18} \\
B (180 days) & 58.49 $\pm$ 5.28 & 29.26 $\pm$ 0.46 & \textbf{2.00 $\pm$ 0.18} \\
        \bottomrule
    \end{tabular}
\end{table}

\section{Discussion}
This work introduces \texttt{TumorTwin}, an open-source python framework for initializing and personalizing image-based DTs for oncology applications. The codebase is modular and adaptable, and represents a significant step towards a common framework  for pan-cancer DTs. Many DT use-cases in oncology, for example, tumor response prediction~\cite{hormuth2021image}, radiotherapy optimization~\cite{Lipkova2019}, and chemotherapy optimization~\cite{Wu2022} are shared across disease sites, so success in one site can be rapidly translated and developed in other sites using this framework. By leveraging a modular architecture, researchers can easily integrate alternative data sources, tumor growth models, treatment models, and numerical algorithms, enabling efficient exploration of modeling choices and their impact on DT performance. This is supported by a complete and robust default architecture, with support for efficient gradient computation and GPU acceleration.

The are several opportunities for further development which can leverage our modular approach to extend the functionality of \texttt{TumorTwin}. The first of these is the introduction of high-dimensional model parameters. Homogenization of model parameters leads to a parsimonious model that can efficiently be optimized, but lacks the ability to faithfully capture the intricate intra-tumoral heterogeneity observed in the real world. High-dimensional parameters have been used by the authors and others to better describe the heterogeneous nature of the tumor and its microenvironment through, for example, spatially-varying proliferation rates~\cite{hormuth2021image}, heterogeneous delivery of chemotherapy~\cite{jarrett_quantitative_2021}, tissue-specific diffusion coefficients~\cite{Swan2018,Lipkova2022}, and tissue mechanical properties~\cite{chen2013}. However, introducing high-dimensional parameters increases computational overhead and will require further developments, including scalable methods and surrogates~\cite{metzcar2024,christenson2024fast}, to enable analysis in clinically actionable timelines. Such models could easily replace the full-order high-fidelity model as a new \texttt{TumorGrowthModel} in our framework.

A second avenue for future development involves uncertainty quantification and subsequent optimization under uncertainty. Despite considerable uncertainties in the data collection process and data-driven estimation of tumor properties, the uncertainty quantification of tumor growth models is still in its infancy~\cite{hawkins2013bayesian,lima2017selection,Lipkova2019,liang2023bayesian,chaudhuri_dt_2023}. In addition to increased certifiability of model predictions, uncertainty quantification opens the door for robust decision-making through the minimization of tail risks~\cite{tyrrell2015engineering,chaudhuri2022certifiable,chaudhuri_dt_2023}. This codebase establishes a high-performance foundation which we plan to extend with rigorous uncertainty quantification to establish trust and accuracy and enable systemic model validation~\cite{lorenzo2025validating}.

A final avenue for future development involves the integration of machine learning techniques into the DT architecture. Since \texttt{TumorTwin} is tightly integrated with Pytorch, it can seamlessly integrate with other pytorch-based machine learning models. For example, a Pytorch-based convolutional neural network may be integrated as part of the MRI processing pipeline, or a neural network could be used to represent a spatially varying model parameter with gradient computation naturally extending to the network hyperparameters.

\section{Methods}
\subsection{Description of Synthetic Cases and Treatment Regimens}
We created two synthetic datasets to demonstrate \texttt{TumorTwin} in two different disease sites and treatment paradigms. In particular, we consider HGG and TNBC, where both tests cases were synthesized using images from publicly available datasets. The HGG example provides synthetic longitudinal collected before,during and after RT, with the RT and CT regimens dictated by the current standard of care treatment protocol~\cite{Stupp2005}. The treatment consists of RT (delivered in 2 Gy fractions over 30 sessions (5 days per week for 6 weeks), alongside daily oral temozolomide at 75 $\text{mg}/\text{m}^2$ for the entire radiotherapy period. Following a 4-week break, patients undergo 6 cycles of adjuvant CT, administered at 150–200 $\text{mg}/\text{m}^2$ per day for 5 days in each 28-day cycle. Likewise, the TNBC example provides synthetic longitudinal data collected before, during and after the delivery of neoadjuvent CT. In the synthetic case, we consider the weekly delivery of Paclitaxel administered at 80 $\text{mg}/\text{m}^2$ weekly for 12 weeks.  In this section we detail the image processing and modeling techniques used to generate both synthetic datasets.

\subsection{Generation of synthetic longitudinal imaging studies}\label{sec:data-generation}
For the HGG case, we seeded an artificial tumor within the SRI24 normal adult brain atlas~\cite{rohlfing2010sri24} and evolved it using Eqs. \eqref{eqn:RD}-\eqref{eqn:RT} with the following parameter values:  $k = 0.05, D = 0.1\ \text{mm}^2/\text{s}, \alpha_\text{RT} = 0.05\ \text{Gy}^{-1}, \beta_\text{RT} = 0.005\ \text{Gy}^{-2},  \alpha_1 = 0.2\ \text{day}^{-1}$,  and $\beta_1 = 9.242\ \text{day}^{-1}$~\cite{newman1996pharmacokinetics}. This atlas also provided $T_1$ and $T_2$ weighted images which we used to define the brain mask. The synthetic HGG growth and response was sampled every 45 days until day 225 and the numerical time step was assigned to $\delta t = 0.1\ \text{day}^{-1}$.

For the TNBC data we used anatomical ($T_1$-weighted pre- and post-contrast) and functional ($ADC$) imaging data from case 104268 from the Investigation of Serial studies to Predict Your Therapeutic Response with Imaging And molecular analysis 2 (I-SPY 2) dataset~\cite{newitt2021acr,li2022ispy2}. All images were registered to the $T_1$-weighted MRI using a rigid registration using a rigid registration using \texttt{imregtform} in MATLAB R2024b~\cite{MATLAB}. We then used the radiologist drawn tumor segmentation included within the I-SPY2 dataset. A breast mask was created by a manual intensity threshold followed by filling holes using \texttt{imfill} in MATLAB.  Normalized tumor cell density maps were generated using Equation~\eqref{eqn:adc}. Using the pre-treatment visit, we then simulated TNBC growth using the following parameters:  $D = 0.1\ \text{mm}^2/\text{s}$, $k = 0.05$, $\alpha_\text{RT} = 0\ \text{Gy}^{-1}$, $\beta_\text{RT} = 0\ \text{Gy}^{-2}$,  $\alpha = 0.2\ \text{day}^{-1}$,  and $\beta = 0.7\ \text{day}^{-1}$~\cite{newman1996pharmacokinetics}. The synthetic TNBC growth and response was sampled at the time points available for the real I-SPY dataset. 

In both cases, NIFTI-formatted images of the tumor segmentations are generated at each visit by applying a threshold of  $N(\bs{x},t) > 0.001$. We then used the segmentations, $N(\bs{x},t)$, and Equation~\eqref{eqn:adc} to calculate the $ADC$ for each corresponding $N(\bs{x},t)$ map which is saved as NIFTI-formatted images. Lastly, we create unique JSON patient configuration files detailing the treatment and imaging schedule for each synthetic case.

\section*{Availability of source code and requirements}
\begin{itemize}
\item Project name: \texttt{TumorTwin}
\item Project home page: \href{https://github.com/OncologyModelingGroup/TumorTwin}{\texttt{TumorTwin GitHub}}
\item Operating system(s): Platform independent
\item Programming language: Python (version: \texttt{">=3.9, <3.12"})
\item Other requirements: \href{https://github.com/OncologyModelingGroup/TumorTwin/blob/main/pyproject.toml}{\texttt{TumorTwin TOML}}
\item License:  \href{https://github.com/OncologyModelingGroup/TumorTwin/blob/main/LICENSE.md}{\texttt{TumorTwin License}}\\
Any restrictions to use by non-academics: \href{https://github.com/OncologyModelingGroup/TumorTwin/blob/main/LICENSE.md}{Commercial use restrictions}
\end{itemize}

\subsection*{Competing Interests}
The authors declare that we have three patents related filed or pending related to the image-based modeling approach employed in this manuscript:  US-20230274842-A1, W02023049207A1, Provisional application 63/495,87. 

\subsection*{Funding}
We acknowledge support for this project via Frederick National Laboratory for Cancer Research Subcontract numbers 23X068Q and 23X068QF1. DAH, EABFL, RB, and TEY acknowledge support from  National Cancer Institute R01CA235800, U24CA226110, U01CA174706, CPRIT RP220225, and IRG-21-135-01-IRG from the American Cancer Society. TEY is a CPRIT Scholar in Cancer Research. AC, DAH, GP, TEY, and KW acknowledge support from the National Science Foundation (NSF) FDT-Biotech award 2436499.

\subsection*{Author's Contributions}
MK contributed to conceptualization, methodology, software--design, software--implementation, software--validation, writing--original draft, writing--review and editing. 
AC contributed to conceptualization, methodology, software, validation, funding acquisition, writing--original draft, writing--review and editing. 
EABF contributed to methodology, software, validation, writing--original draft, writing--review and editing. 
RB contributed to methodology, software, validation, writing--original draft. 
GP contributed to methodology, software, validation, writing--original draft, writing--review and editing. 
KW contributed to conceptualization, project administration, funding acquisition, supervision writing--original draft, writing--review and editing. 
TEY contributed to conceptualization, project administration, funding acquisition, supervision writing--original draft, writing--review and editing. 
DAH contributed to conceptualization, methodology, resources, software, validation, funding acquisition, writing--original draft, writing--review and editing.

\subsection*{Data availability}
The data sets supporting the results of this article are available in the code repository under \texttt{input\_files}. \url{https://github.com/OncologyModelingGroup/TumorTwin}.

\section*{Declarations}
\subsection*{List of abbreviations}
CPU: central processing unit, CT: chemotherapy, DT: digital twin, GPU: graphical processing unit, HGG: high grade glioma, JSON: JavaScript object notation, MRI: magnetic resonance imaging, NIFTI: neuroimaging informatics technology initiative, ODE: ordinary differential equations, RT: radiotherapy TNBC: triple-negative breast cancer,

\subsection*{Ethical Approval}
Not applicable

\subsection*{Consent for publication}
Not applicable

\appendix
\section{Appendix A - Model prediction and calibration results for triple-negative breast cancer}\label{sec:appendix}
\begin{figure}
    \centering
    \includegraphics[width=\figsize]{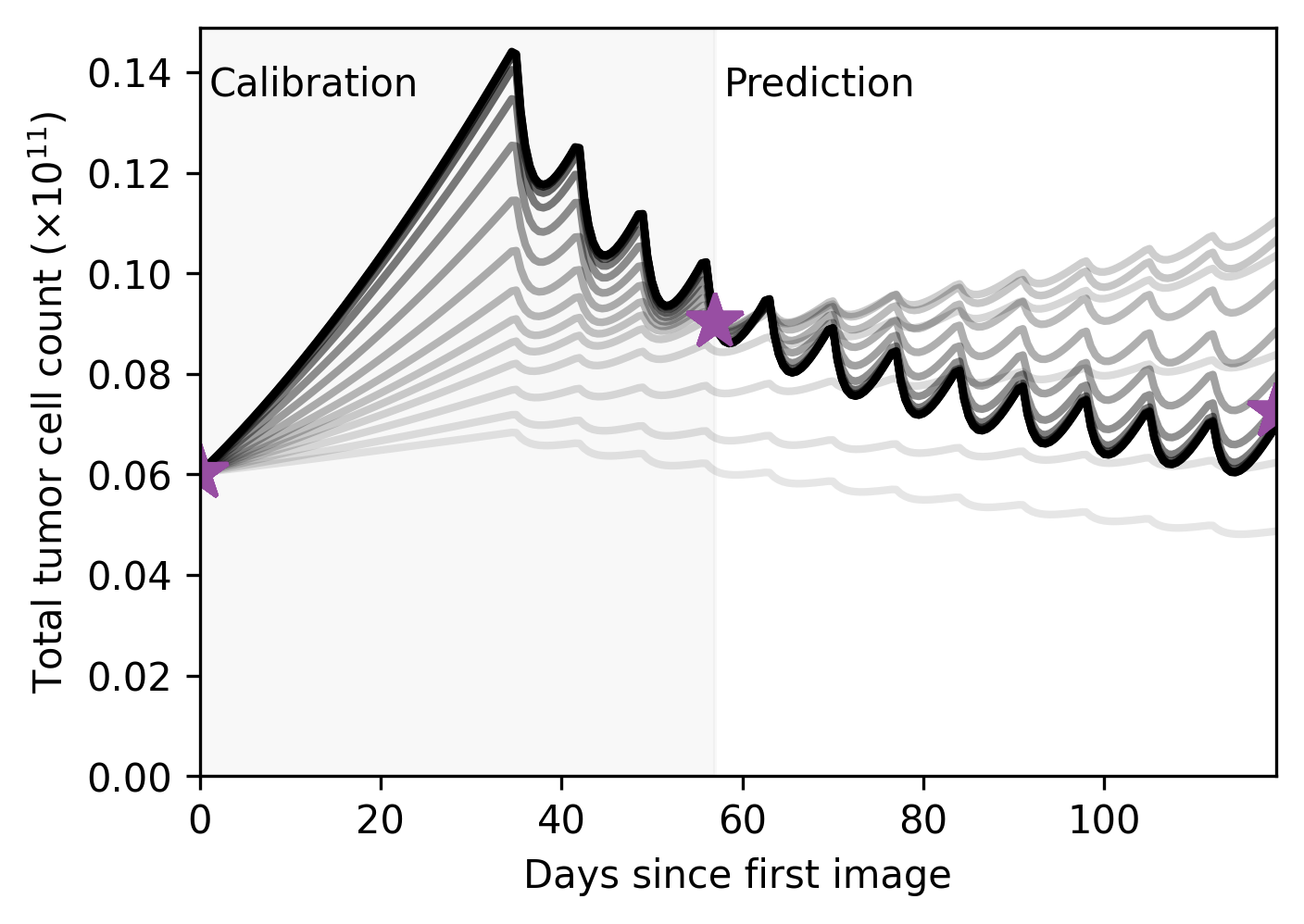}\\ \vspace{-0.5em}
    \includegraphics[width=\figsize]{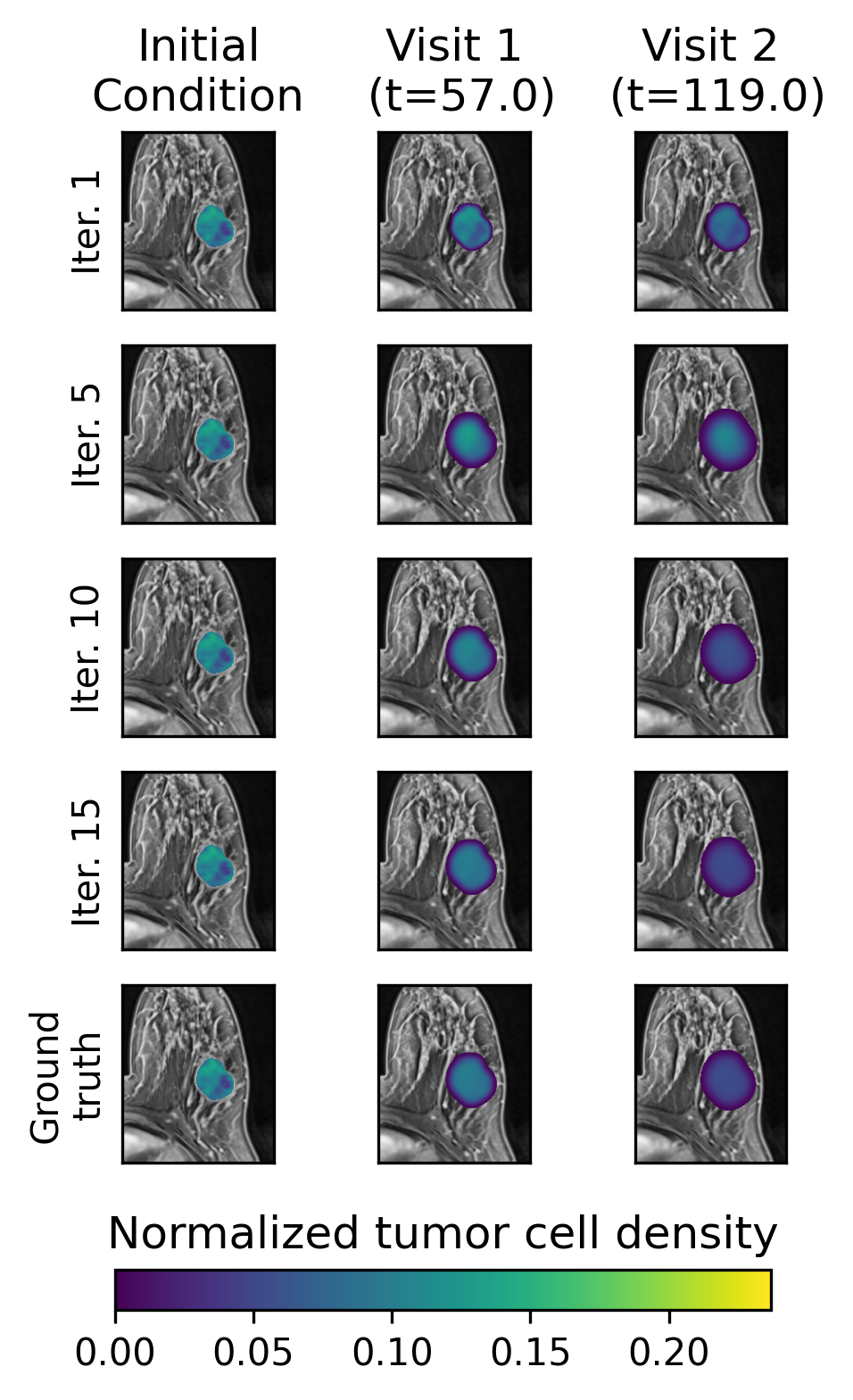}\\  \vspace{-0.5em}
    \includegraphics[width=\figsize]{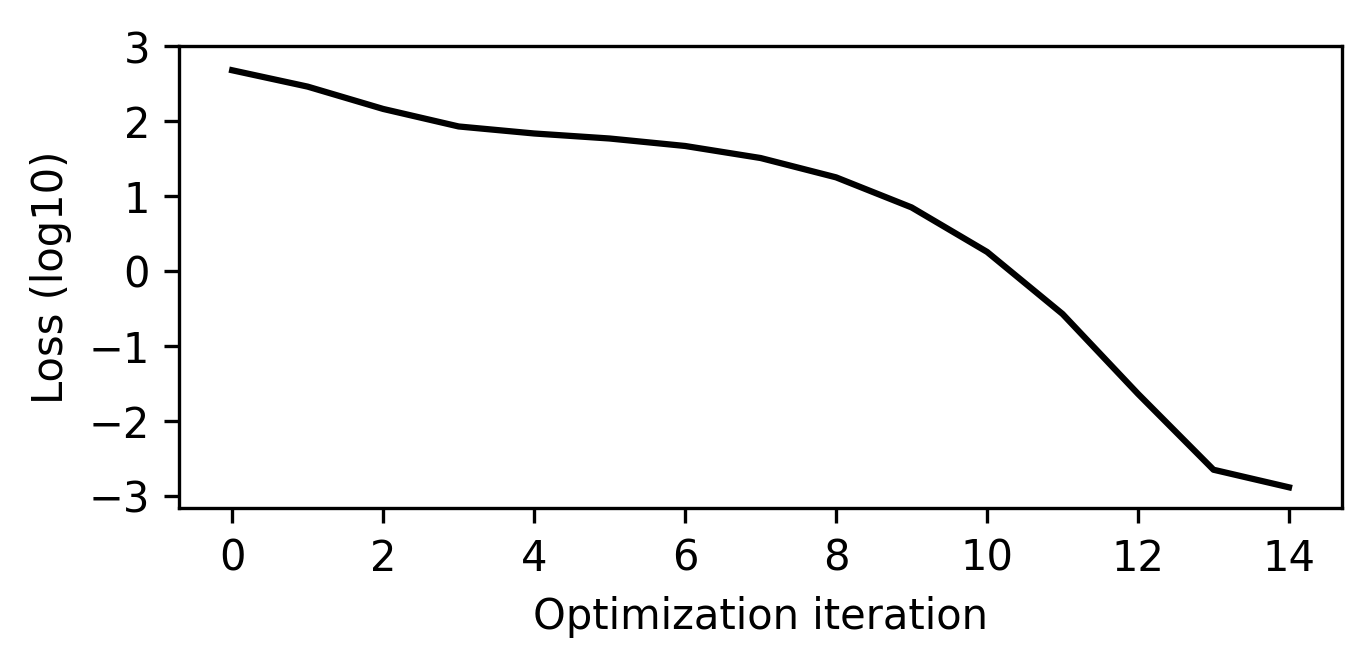}    
    \caption{\textit{Model calibration to patient-specific MRI data - Triple-negative breast cancer example}. Top: Total tumor cell count (TTC) time series for each iteration of the calibration (black lines; opacity increased with iteration) compared with the observed TTC (purple stars). Middle: Evolution of a central tumor slice across time (left-to-right), and across optimization iterations (top-to-bottom). The bottom row shows the ground-truth data used for calibration and validation. Bottom: Convergence of the loss function over optimization iterations.}
    \label{fig:calibration_tnbc}
\end{figure}

To demonstrate the applicability of the \texttt{TumorTwin} framework to different cancer sites, we here present calibration results for the \textit{in-silico} TNBC dataset described in the Methods section.  Figure \ref{fig:calibration_tnbc} presents these results, and is analogous to the calibration results for HGG provided in \ref{fig:calibration}. Here we again use the LM optimizer to calibrate a patient-specific DT model to the first two visits of MRI data, leaving the third visit out to assess predictive capability. For this example, we know the ground-truth model parameters (used to generate the dataset), but the optimizer is initiated with an initial guess of $k=0.005$, $D=0.01$, and $\alpha_1=0.04$ (20\% of the truth values). Other parameters are fixed to their ground truth values. The optimizer parameters are the same default values as used for the HGG demonstration. We again observe that the optimizer is able to calibrate the unknown model parameters to match the input MRI data within a few iterations, and is again robust to large variations in the solution across iterations. 

\bibliographystyle{abbrv}
\bibliography{paper-refs}

\end{document}